\documentclass[12pt]{elsart}
\usepackage{amstex}
\catcode`\@=11
\def\@journal{{\tt hep-th/9609113}}
\catcode`\@=12
\renewcommand{\d}{\partial}
\newcommand{\A}{{\ensuremath{\mathsf A}}}
\newcommand{\B}{{\ensuremath{\mathsf B}}}
\newcommand{\C}{{\ensuremath{\mathsf C}}}
\newcommand{\D}{{\ensuremath{\mathsf D}}}
\newcommand{\G}{{\ensuremath{\mathsf G}}}
\newcommand{\J}{{\ensuremath{\mathsf J}}}
\newcommand{\K}{{\ensuremath{\mathsf K}}}
\newcommand{\M}{{\ensuremath{\mathsf M}}}
\newcommand{\N}{{\ensuremath{\mathsf N}}}
\renewcommand{\P}{{\ensuremath{\mathsf P}}}
\newcommand{\R}{{\ensuremath{\mathsf R}}}
\newcommand{\T}{{\ensuremath{\mathsf T}}}
\newcommand{\W}{{\ensuremath{\mathsf W}}}
\newcommand{\X}{{\ensuremath{\mathsf X}}}

\newcommand{\bC}{{\ensuremath{\mathbb C}}}
\newcommand{\bZ}{{\ensuremath{\mathbb Z}}}
\newcommand{\NPB}[3]{{\sl Nucl. Phys.\/} {\bf B#1} (#2) #3}
\newcommand{\CMP}[3]{{\sl Comm. Math. Phys.\/} {\bf #1} (#2) #3}
\newcommand{\PLB}[3]{{\sl Phys. Lett.\/} {\bf #1B} (#2) #3}
\newcommand{\JDG}[3]{{\sl J. Diff. Geometry\/} {\bf #1} (#2) #3}
\newcommand{\Invm}[3]{{\sl Invent. math.\/} {\bf #1} (#2) #3}
\newcommand{\AGAG}[3]{{\sl Ann. Global Anal. Geom.\/} {\bf #1} (#2) #3}

\newcommand{\IJMPA}[3]{{\sl Int. J. Mod. Phys.\/} {\bf A#1} (#2) #3}
\newcommand{\IJMPC}[3]{{\sl Int. J. Mod. Phys.\/} {\bf C#1} (#2) #3}
\newcommand{\MPLA}[3]{{\sl Mod. Phys. Lett.\/} {\bf A#1} (#2) #3}
\newcommand{\SM}[3]{{\sl Selecta Math.\/} {\bf A#1} (#2) #3}
\begin{document}
\begin{frontmatter}
\title{A note on the extended superconformal algebras associated with
manifolds of exceptional holonomy}
\author[QMW]{Jos\'e~M~Figueroa-O'Farrill\thanksref{EPSRC}}
\address[QMW]{$\langle\text{\tt j.m.figueroa@qmw.ac.uk}\rangle$\\
Department of Physics, Queen Mary and Westfield College,
Mile End Road, London E1 4NS, UK}
\thanks[EPSRC]{Supported by the EPSRC under contract GR/K57824.}
\begin{abstract}
It was observed some time ago by Shatashvili and Vafa that superstring
compactification on manifolds of exceptional holonomy gives rise to
superconformal field theories with extended chiral algebras.  In their
paper, free field realisations are given of these extended
superconformal algebras inspired by Joyce's constructions of such
manifolds as desingularised toroidal orbifolds.  The purpose of this
note is to give another realisation of these algebras starting not
from free fields, but from the superconformal algebras associated to
Calabi--Yau manifolds.  These superconformal algebras, originally
studied by Odake, are extensions of the $N{=}2$ Virasoro algebra.
For the case of $G_2$ holonomy, our realisation is inspired in the
conjectured construction of such manifolds as a desingularisation of
$(K \times S^1)/\bZ_2$, where $K$ is a Calabi--Yau 3-fold admitting an
antiholomorphic involution.  Similarly, for the case of $Spin(7)$
holonomy our realisation suggests a construction of such manifolds as
desingularisations of $K'/\bZ_2$, where $K'$ is a Calabi-Yau 4-fold
admitting an antiholomorphic involution.
\end{abstract}
\end{frontmatter}

\section{Introduction}

In the context of M-theory \cite{Duff} and F-theory \cite{Vafa},
compactification to four dimensions requires that we do so on
manifolds of seven and eight dimensions, respectively.  If we require
supersymmetry in four-dimensions we are forced to compactify on
manifolds admitting parallel spinors.  This in turn constraints the
holonomy group of the manifold to be contained in the isotropy group
of the spinor.  For $M$ an irreducible riemannian $n$-manifold which
is not locally symmetric, the possible holonomy groups are those in
Berger's list.  Of the groups in that list, only $SU(n/2)$, $Sp(n/4)$,
$G_2$ (for $n{=}7$) and $Spin(7)$ (for $n{=}8$) admit parallel spinors
(see, for example, \cite{Wang}).  For $M$ a 7-dimensional
simply-connected manifold to admit the minimum (nonzero) number of
parallel spinors, yielding the minimum number of four-dimensional
supersymmetries, its holonomy group must be $G_2$.  The same
conditions on an 8-dimensional manifold singles out those with
$Spin(7)$ holonomy.  In both of these cases there is only one parallel
spinor.

For traditional superstring compactification from ten to four
dimensions, the desired holonomy is $SU(3)$.  Thanks to Yau's solution
of the Calabi conjecture, any K\"ahler $n$-fold with vanishing
canonical class admits a unique metric of $SU(n)$ holonomy in the same
K\"ahler class.  This provides us with many examples of such
manifolds.  On the other hand, relatively few examples are known of
compact manifolds of exceptional holonomy.  It was not until two years
ago that the first such manifolds were constructed by Joyce
\cite{JoyceG2,JoyceSpin7} by desingularising toroidal orbifolds.

Such manifolds are not just interesting in their own right, but their
study is relevant for superstring phenomenology.  Just as in the case
of Calabi--Yau 3-folds before them, it is hoped that much can be
learned by exploring the superconformal field theories they give rise
to.  A particularly fruitful approach is to study the orbifold limit,
since although the geometry becomes singular, the conformal field
theory does not.  Moreover, it seems that the orbifold limit captures
some of the information on the desingularisation process \cite{SV};
for example, different desingularisations seem to be correspond to the
different conformal field theories associated with the same orbifold
via the process of turning on discrete torsion \cite{VafaDT}.  This
prompted the {\em generalised mirror conjecture\/} in \cite{SV}, for
which more evidence would be welcome.

The superconformal algebras arising from compactification on
Calabi--Yau $n$-folds are well known \cite{Odake}.  They correspond to
extensions of the $N{=}2$ Virasoro algebra by a complex field of
dimension $n/2$.  For $n{=}1$ it is an extension of the $N{=}2$
Virasoro algebra by a complex fermion and a complex boson, whereas for
$n{=}2$ it becomes the (small) $N{=}4$ Virasoro algebra.  This is
expected since $SU(2) = Sp(1)$ and all 4-manifolds with $Sp(1)$
holonomy are hyperk\"ahler.  These algebras exist for generic values
of the Virasoro central charge.  For $n{=}3$ and $n{=}4$, the cases of
interest in the present note, the Virasoro central charge is fixed to
9 and 12 respectively, corresponding to compactification manifolds of
dimensions 6 and 8, respectively.

For compactifications on manifolds of exceptional holonomy, the
resulting superconformal algebras are extensions of the $N{=}1$
Virasoro algebra by superfields of weights $\tfrac{3}{2}$ and $2$ for
$G_2$, and $2$ for $Spin(7)$.  These algebras were written down by
Shatashvili and Vafa in \cite{SV} for the first time in the present
context; although classical versions of these algebras had appeared in
\cite{HP}. The $Spin(7)$ algebra had been discovered previously in a
different context \cite{FS}.\footnote{The $Spin(7)$ algebra was also
written down in \cite{BEHH}, where the $G_2$ algebra appears for the
first time.  The $G_2$ algebra had also appeared in \cite{Mallwitz}. I
thank Andreas Honecker for reminding me of this work.}  The $G_2$
algebra exists only for $c=\frac{21}{2}$, corresponding to
7-dimensional compactification manifolds, whereas the $Spin(7)$
algebra belongs to a one-parameter family of superconformal algebras
\cite{FS}, of which the $c{=}12$ point is the interesting one in the
present context.

The algebras in \cite{SV} were constructed in a free field realisation
appropriate to the study of those manifolds which are presented as
desingularised toroidal orbifolds; however other constructions may
exist and it is desirable to understand them in the language of
conformal field theory.  The purpose of this note is to construct new
realisations of this algebra which belie constructions of these
manifolds starting from Calabi--Yau 3- and 4-folds.

This note is organised as follows.  In Section 2 we describe the
extended superconformal algebras of interest: the ones associated to
manifolds of $SU(3)$, $SU(4)$, $G_2$ and $Spin(7)$ holonomies.  In
\cite{JoyceG2} a construction of compact manifolds with $G_2$ holonomy
is conjectured, which consists in desingularising an orbifold $(K
\times S^1)/\bZ_2$, where $K$ is a Calabi--Yau 3-fold admitting an
antiholomorphic involution, and the generator $\sigma$ of the $\bZ_2$
acts as the involution on $K$, and as inversion on the circle.  This
suggests that there should be a realisation of the $G_2$
superconformal algebra in terms of the superconformal algebra
associated to the Calabi--Yau 3-fold $K$, and to the circle.  Roughly
the geometric $\bZ_2$ induces an automorphism of the superconformal
algebra and inside the fixed subalgebra one finds a realisation of the
algebra in \cite{SV}.  This will shown in section 3.  There we also
show that if we take a Calabi--Yau 4-fold $K'$ admitting an
antiholomorphic involution $\sigma$, then the conformal field theory
associated to the orbifold $K'/\langle\sigma\rangle$ embeds a
superconformal subalgebra isomorphic to the $Spin(7)$ algebra.  This
suggests a construction of compact 8-manifolds with $Spin(7)$ holonomy
obtained by desingularising the orbifold.  After completion of this
work, we became aware of a preprint \cite{BBMOOY} which mentions the
large volume limit of this last embedding.

Throughout we use the notation $[A,B]_\ell$ to denote the residue of
the $\ell$-th order pole in the operator product expansion of the
fields $A$ and $B$:
\begin{equation*}
A(z) B(w) = \sum_{\ell\ll \infty} \frac{[A,B]_\ell(w)}{(z-w)^\ell}~,
\end{equation*}
and assume familiarity with the axiomatics of these brackets as
explained, for example, in \cite{Kris}.

\section{The superconformal algebras}

In this section we write down the superconformal algebras associated
with compactifications on manifolds of holonomy $SU(3)$ and $SU(4)$
\cite{Odake}, and $G_2$ and $Spin(7)$ \cite{SV}.  Let $M$ be an
irreducible manifold with holonomy group in the above list.  Every
parallel form on $M$ gives rise to a generator of the algebra, by
pulling the form back to the worldsheet using the fermions.  Parallel
forms are precisely the singlets under the holonomy group in the
representation $\bigwedge^*T$ where $T$ is the irreducible
representation on tangent vectors.  Hence group theory alone tells us
all about the parallel forms (indeed tensors) on $M$; or
alternatively we can use the parallel fermions to construct these
forms as bispinors.

\subsection{The superconformal algebra of a Calabi--Yau $n$-fold}
\label{sec:CY}

For $SU(n)$ holonomy, the tangent vectors form an irreducible
$2n$-dimensional representation $T$.  Its complexification splits into
$T_\bC = T' \oplus T''$, where $T'$ is the fundamental complex
representation of $SU(n)$ and $T''$ is its complex conjugate.  The
parallel forms are in one-to-one correspondence with the $SU(n)$
singlets in $\bigwedge^* T$.  These include the K\"ahler form $\omega
\in T'\otimes T'' \subset \bigwedge^2 T$ and its powers, but also the
real 2-dimensional representation $\bigwedge^n T' \oplus \bigwedge^n
T''$, corresponding to the real and imaginary parts of a complex
$(n,0)$-form $\Omega$.  In the large volume limit, we can write down
the following generators relative to a complex coordinate basis:
\begin{align}
J = & \half \omega_{a\bar b} \psi^a\psi^{\bar b}\notag\\
H = & \tfrac{1}{n!} \Omega_{a_1a_2\cdots a_n} \psi^{a_1}\psi^{a_2}\cdots
\psi^{a_n}\label{eq:fields}\\
\bar H = & \tfrac{1}{n!} \bar\Omega_{\bar a_1\bar a_2\cdots \bar a_n}
\psi^{\bar a_1}\psi^{\bar a_2}\cdots \psi^{\bar a_n}~,\notag
\end{align}
which will define an extension of the $N{=}1$ Virasoro algebra written
for the first time in \cite{Odake}.

Let us now write down these algebras.  We let $\T$ and $\G$ denote the
generators of the $N{=}1$ Virasoro algebra with central charge $3n$.
We let $\J$ be a weight one superconformal primary normalised to
$[\J,\J]_2 = -n$.  Together with its superpartner $\G' = [\G,\J]_1$,
they generate the $N{=}2$ Virasoro algebra.  Now let $\A$ and $\B$ be
$N{=}1$ superconformal primaries of weight $n/2$.  They are to be
understood as the generators corresponding to the real and imaginary
parts of the field $H$ in \eqref{eq:fields}.  They satisfy the
following operator product expansion with $\J$:
\begin{alignat*}{2}
[\J,\A]_1 &= -n \B &\quad\text{and}\quad [\J,\B]_1&= n \A~.
\end{alignat*}
We let $\C = [\G,\A]_1$ and $\D = [\G,\B]_1$ denote their
superpartners.  Because $\Omega$ is actually antiholomorphic,
$\d\Omega = 0$, $(\A,\C)$ and $(\B,\D)$ are $N{=}2$ (anti)chiral
superfields.  This means that they obey the following operator product
expansions with the second supersymmetry generator: $[\G',\A]_1 = -\D$
and $[\G',\B]_1= \C$.  The remaining operator product expansions are
given in terms of the ones involving the primary fields $\A$ and $\B$,
so we give only these.  The others can be reconstructed using the
Jacobi-like identities of the $[-,-]_\ell$ brackets, or equivalently
the associativity of the operator product expansion.  It is here that
we must distinguish between $n{=}3$ and $n{=}4$.

\subsubsection{$n{=}3$}
\label{sec:su3}

The following operator product expansions hold:
\begin{xalignat*}{2}
[\A,\A]_3 &= -4 & \quad [\A,\A]_1 &= 2 (\J\J)\\
[\B,\B]_3 &= -4 & \quad [\B,\B]_1 &= 2 (\J\J)\\
[\A,\B]_2 &= -4\J & \quad [\A,\B]_1 &= -2 \d\J\\
[\A,\C]_2 &= -2 \G & \quad [\A,\C]_1 &= -2 (\J\G')\\
[\A,\D]_2 &= -2 \G' & \quad [\A,\D]_1 &= -2 (\J\G)\\
[\B,\C]_2 &= -2 \G' & \quad [\B,\C]_1 &= 2 (\J\G)\\
[\B,\D]_2 &= 2 \G & \quad [\B,\D]_1 &= 2 (\J\G')~,
\end{xalignat*}
where the normal-ordered product $(AB)$ is defined by $(AB) =
[A,B]_0$.  As it stands, the chiral algebra defined by these brackets
is not associative.  The Jacobi-like identities are only satisfied
modulo the ideal generated by the weight $\tfrac{5}{2}$ fields
\begin{equation}
\N^{(1)}_{\mathrm{CY}} = \d\A - (\J\B)\qquad\text{and}\qquad 
\N^{(2)}_{\mathrm{CY}} = \d\B + (\J\A)~.\label{eq:idealCY}
\end{equation}
These fields are null for this value of the central charge $c{=}9$.

\subsubsection{$n{=}4$}
\label{sec:su4}

Similarly in this case, the following operator product expansions
hold:
\begin{alignat*}{3}
[\A,\A]_4 &= -8 &\quad [\A,\A]_2 &= -4 (\J\J) &\quad [\A,\A]_1 &= -4
(\d\J\J)\\
[\B,\B]_4 &= -8 &\quad [\B,\B]_2 &= -4 (\J\J) &\quad [\B,\B]_1 &= -4
(\d\J\J)\\
[\A,\B]_3 &= -8\J &\quad [\A,\B]_2 &= 4 \d\J &\quad [\A,\B]_1 &=
-\tfrac{4}{3} (\J\J\J) + \tfrac{4}{3}\d^2\J\\
[\A,\C]_3 &= -4\G &\quad [\A,\C]_2 &= -4 (\J\G') & \quad [\A,\C]_1 &=
2 (\G\J\J) - 2 (\d\J\G')\\
[\A,\D]_3 &= 4\G' &\quad [\A,\D]_2 &= -4 (\J\G) & \quad [\A,\D]_1 &=
-2 (\J\J\G') - 2 (\d\J\G)\\
[\B,\C]_3 &= -4\G' &\quad [\B,\C]_2 &= 4 (\J\G) & \quad [\B,\C]_1 &= 2
(\J\J\G') + 2 (\d\J\G)\\
[\B,\D]_3 &= -4\G &\quad [\B,\D]_2 &= -4 (\J\G') & \quad [\B,\D]_1 &=
2 (\G\J\J) - 2 (\d\J\G')~,
\end{alignat*}
where the normal-ordered product associates to the left; that is,
$(ABC) = (A(BC))$.  Again the above operator product expansions are
associative only modulo the ideal generated by the fields given in
\eqref{eq:idealCY}, which now have weight 3 and are null for
$c{=}12$.

We should remark that it follows from the above brackets that the
superconformal algebras associated with a Calabi--Yau $n$-fold have an
additional automorphism, corresponding to multiplying the complex
fields $\A + i \B$ and $\C + i \D$ by the same phase.

\subsection{The algebra associated with $G_2$ holonomy}
\label{sec:g2}

In a 7-dimensional irreducible manifold of $G_2$ holonomy, the tangent
vectors are in an irreducible representation $T$ of $G_2$.  Computing
the singlets in $\bigwedge^* T$, we find that there is a unique
parallel 3-form $\phi$, a unique parallel 4-form $\star \phi$, and
their product.  In the large volume limit we can write the fields
\begin{align*}
\Phi &= \tfrac{1}{3!} \phi_{ijk} \psi^i\psi^j\psi^k\\
\Phi^* &= \tfrac{1}{4!} (\star\phi)_{ijkl} \psi^i\psi^j\psi^k\psi^l~,
\end{align*}
which generate an extension of the $N{=}1$ Virasoro algebra.

Let us now write down this algebra.  We let $\T$ and $\G$ be the
generators of the $N{=}1$ Virasoro algebra with $c{=}\frac{21}{2}$.
Let $\P$ be a superprimary field of weight $\frac{3}{2}$ with operator
product expansion $[\P,\P]_3  = -7$ and $[\P,\P]_1 = 6 \X$, which
defines $\X$.  The field $\X$ has weight 2, but it is is not a primary
since $[\T,\X]_3  = -\tfrac{7}{4}$.  In addition, we have
\begin{alignat*}{3}
&&[\P,\X]_2 & = -\tfrac{15}{2} \P &\quad [\P,\X]_1 &= -\tfrac{5}{2}
\d\P~,\\
\intertext{and}
[\X,\X]_4 & = \tfrac{35}{4} &\quad [\X,\X]_2 &= -10\X &\quad [\X,\X]_1
&= -5 \d\X~.
\end{alignat*}
It follows from these formulae that $\G' \equiv \pm
\tfrac{i}{\sqrt{15}} \P$, and $\T' \equiv -\tfrac{1}{5} \X$ satisfy
an $N{=}1$ Virasoro algebra with central charge $c'{=}\tfrac{7}{10}$
corresponding to the tricritical Ising model.  We now define $\K =
[\G,\P]_1$ and $\M=[\G,\X]_1$ as the superpartners of $\P$ and $\X$
respectively.  Because $\P$ is a superconformal primary, $\K$ is a
Virasoro primary of weight $2$.  On the other hand $\M$ is not a
primary because neither is $\X$.  Instead we have $[\T,\M]_3 =
[\G,\X]_2 = -\half \G$ and in addition:
\begin{alignat*}{3}
[\G,\M]_4 & = -\tfrac{7}{2}  &\quad [\G,\M]_2&= \T + 4\X &\quad
[\G,\M]_1 &= \d\X~.\\
\intertext{The rest of the relevant operator product expansions are:}
&&[\P,\K]_2 & = -3 \G &\quad [\P,\K]_1 &= -3 \M - \tfrac{3}{2}\d\G\\
&&[\P,\M]_2 & = \tfrac{9}{2}\K &\quad [\P,\M]_1 &= 3 (\P\G) - \half
\d\K\\
&&[\X,\K]_2 & = -3\K &\quad [\X,\K]_1 &= -3 (\P\G)\\
\intertext{and}
[\X,\M]_3 & = -\tfrac{9}{2}\G &\quad [\X,\M]_2 &= -5\M -
\tfrac{9}{4}\d\G &\quad [\X,\M]_1 &= 4 (\X\G) + \half \d\M +
\tfrac{1}{4} \d^2\G~.
\end{alignat*}
We can obtain the remaining operator product expansions by using the
associativity axiom.  For example, to compute $[\K,\M]_p$ we use the
fact that $\K = [\G,\P]_1$ and that $[\G,-]_1$ is an odd derivation
over all the brackets:
\begin{align*}
[\K,\M]_p &= [[\G,\P]_1,\M]_p\\
          &= [\G,[\P,\M]_p]_1 + [\P,[\G,\M]_1]_p\\
          &= [\G,[\P,\M]_p]_1 + [\P,\d\X]_p\\
          &= [\G,[\P,\M]_p]_1 + (p-1) [\P,\X]_{p-1} + \d[\P,\X]_p~;
\end{align*}
whence from the above formulae we find:
\begin{alignat*}{3}
[\K,\M]_3 &= -15\P &\quad [\K,\M]_2 &= -\tfrac{11}{2} \d\P &\quad 
[\K,\M]_1 &= 3 (\G\K) - 6 (\T\P)~,
\end{alignat*}
which corrects a typo in equation (1.8) in the first appendix of
\cite{SV}.  (The notation is the same as in \cite{SV} except that here
we call $\P$ what they call $\Phi$, and aside from the above typo, we
are in perfect agreement with their results.)

In \cite{SV}, this algebra was obtained in a free field representation
in terms of seven free bosons and seven free fermions.  As a
consequence, associativity of the operator product expansion is
guaranteed.  Abstractly, however, the Jacobi-like identities in the
above algebra are only satisfied modulo the ideal generated by the
weight $\frac{7}{2}$ null field $\N$ defined by
\begin{equation}
\N = 4 (\G\X) - 2 (\P\K) - 4 \d\M - \d^2\G~,\label{eq:nullG2}
\end{equation}
and vanishing identically in the free field realisation of \cite{SV}.

\subsection{The algebra associated with $Spin(7)$ holonomy}
\label{sec:spin7}

We finally look at the case of $Spin(7)$ holonomy.  Let $M$ be an
irreducible 8-dimensional manifold with $Spin(7)$ holonomy.  The
tangent vectors are in the spinorial representation of $Spin(7)$, which
in this context we call $T$.  The only singlets in $\bigwedge^* T$ are
a self-dual 4-form $\Theta$ and its square.  In the large volume
limit, $\Theta$ gives rise to a weight $2$ field:
\begin{equation*}
X = \tfrac{1}{4!} \Theta_{ijkl}\psi^i\psi^j\psi^k\psi^l~,
\end{equation*}
which generates an extension of the $N{=}1$ Virasoro algebra.

Again let $\T$ and $\G$ denote the generators of the $N{=}1$ Virasoro
algebra with central charge $c{=}12$.  We now let $\X$ be a weight $2$
field which we will choose {\em not\/} to be primary, and $\M =
[\G,\X]_1$ be its superpartner of weight $\tfrac{5}{2}$ but also not
primary.  The following operator product expansions define the
algebra:
\begin{alignat*}{3}
&&[\G,\X]_2 & = \half \G  &\quad [\G,\X]_1 &= \M\\
[\X,\X]_4 & = 16 &\quad [\X,\X]_2 & = 16 \X & \quad [\X,\X]_1 &=
8\d\X\\
[\G,\M]_4 & = 4 &\quad [\G,\M]_2 &= -\T + 4\X &\quad [\G,\M]_1 &=
\d\X\\
[\X,\M]_3 & = -\tfrac{15}{2} \G &\quad [\X,\M]_2 &= 8\M -
\tfrac{15}{4}\d\G &\quad [\X,\M]_1 &= -6 (\G\X) + \tfrac{11}{2} \d\M -
\tfrac{5}{4}\d^2\G~.
\end{alignat*}
Again we can compute all other operator products from these by
associativity.  Notice that the field $\T' = \frac{1}{8} \X$ obeys a
Virasoro algebra with central charge $c'{=}\half$, corresponding to
the Ising model.  In counterpoint to the $G_2$ algebra, this one obeys
associativity abstractly and not modulo an ideal.  The reason is that
the algebra admits a one-parameter (the central charge) deformation
with the same fields.  To see this we simply change basis to primary
fields $\Tilde\X = \X - \frac{1}{3} \T$ and $\Tilde\M = \M -
\frac{1}{6} \d\G$.  Then $\Tilde\X$ is a superconformal primary of
weight $2$, and by the results of \cite{FS} there exists a unique such
extension of the $N{=}1$ Virasoro algebra, which exists for generic
values of the central charge.  In fact, if we further rescale
$\Tilde\X$ to $\Hat\X = \pm \frac{3}{\sqrt{23}} \Tilde\X$, and define
$\Hat\M = [\G,\Hat\X]_1$ as its superpartner, then the new algebra
satisfied by $\{\T,\G,\Hat\X,\Hat\M\}$ agrees with the one in the
appendix of the second reference in \cite{FS} for $c{=}12$.

\section{New realisations}

In this section we come to the main results of this note.  We will
construct new realisations of the $G_2$ and $Spin(7)$ superconformal
algebras in sections \ref{sec:g2} and \ref{sec:spin7} in terms of the
algebras of sections \ref{sec:su3} and \ref{sec:su4}, respectively.

Let $K$ be a Calabi--Yau $n$-fold admitting an antiholomorphic
involution $\sigma$.  Then on the K\"ahler form $\omega$ and the
(anti)holomorphic $n$-forms $\Omega$ and $\Bar\Omega$, we have
$\sigma^* \omega = - \omega$ and $\sigma^* \Omega = - \Bar\Omega$.  At
the level of the algebra associated to such a manifold, the involution
is represented by an automorphism which fixes $\T$, $\G$, $\A$ and
$\C$, and changes the sign of $\J$, $\G'$, $\B$ and $\D$.  It is easy
to check that this is an automorphism of the corresponding
superconformal algebras.  As a result, the subspace of the chiral
algebra fixed by this automorphism will be a subalgebra.

In the case $n{=}4$ we see that this subalgebra contains the $Spin(7)$
superconformal algebra, generated by $\T$ and $\G$ together with
\begin{equation*}
\X\equiv \A - \half (\J\J)\qquad\text{and}\qquad \M \equiv \C -
(\J\G') + \half \d\G~.
\end{equation*}
All the brackets in section \ref{sec:spin7} are obeyed on the nose,
except for $[\X,\M]_1$ (and hence $[\M,\M]_1$) which are obeyed only
modulo the ideal generated by the null fields in \eqref{eq:idealCY}.
Indeed, if we define $\R \equiv [\X,\M]_1 + 6 (\G\X) - \tfrac{11}{2} \d\M
+ \tfrac{5}{4}\d^2\G$, we get that
\begin{align*}
\R &= 6 (\G\A) - 4 (\G'\B) + 2 (\J\D) - 2 \d\C\\
   &= 4    [\G,\N^{(1)}_{\mathrm{CY}}]_1 - 6
[\G',\N^{(2)}_{\mathrm{CY}}]_1~.
\end{align*}

In the case $n{=}3$ we need an auxiliary boson-fermion pair $(j,\psi)$
normalised to $[j,j]_2 = 1$ and $[\psi,\psi]_1=1$ corresponding to the
circle.  This algebra has an automorphism given by $(j,\psi) \mapsto
(-j,-\psi)$, which fixes the generators of an $N{=}1$ Virasoro algebra
with central charge $\tfrac{3}{2}$:
\begin{align*}
\T_{S^1} &\equiv \half (jj) + \half (\d\psi\psi)\\
\G_{S^1} &\equiv (j\psi)~.
\end{align*}
Let $K$ be a Calabi--Yau 3-fold admitting an antiholomorphic
involution $\sigma'$, and let $\sigma$ be the involution on $K\times
S^1$ acting by $\sigma(z,\theta) = (\sigma'(z), -\theta)$.  On the
superconformal algebra corresponding to $K\times S^1$, generated by
$\T_{\mathrm{CY}}$, $\G_{\mathrm{CY}}$, $\J$, $\G'$, $\A$, $\B$, $\C$,
$\D$, $j$ and $\psi$, the involution is represented by the
automorphism which fixes $\T_{\mathrm{CY}}$, $\G_{\mathrm{CY}}$, $\A$
and $\C$, and changes the signs of the other generators.  If we define
\begin{equation*}
\T \equiv \T_{\mathrm{CY}} + \T_{S^1}\qquad
\G \equiv \G_{\mathrm{CY}} + \G_{S^1}\qquad\text{and}\qquad
\P \equiv \A + (\J\psi)~,
\end{equation*}
then the other fields of the $G_2$ algebra follow:
\begin{align*}
\X &\equiv (\B\psi) + \half (\J\J) - \half (\d\psi\psi)\\
\K &\equiv \C + (\J j) + (\G'\psi)\\
\M &\equiv (\D\psi) - (\B j) + (j\d\psi) + (\J\G') - \half \d \G~;
\end{align*}
and computing their brackets, we find those of the $G_2$
superconformal algebra in section \ref{sec:g2}, modulo the ideal
generated by the fields in \eqref{eq:idealCY}.  Also, as expected, the
null field $\N$ in \eqref{eq:nullG2}, belongs to the ideal generated
by the fields in \eqref{eq:idealCY}.

The above embeddings are not unique, of course, since we can still
perform the automorphism mentioned at the end of section \ref{sec:CY},
namely $\A + i \B \mapsto e^{i\theta} (\A + i \B)$ and similarly with
$\C +i\D$.

\begin{ack}
It is a pleasure to thank Bobby Acharya for introducing me to the work
of Joyce, and along with Bas Peeters for many useful conversations on
this and other topics.  I have benefited from the {\sl Mathematica\/}
package {\tt OPEdefs} \cite{OPEdefs}, and it is a pleasure to thank
Kris Thielemans once again for writing it.
\end{ack}

\end{document}